\documentclass[a4paper,11pt,english]{scrartcl}
\usepackage[utf8]{inputenc}
\usepackage[T1]{fontenc}
\usepackage[left=2.5cm,right=3cm,top=2cm,bottom=3.5cm]{geometry}
\usepackage{lmodern}
\usepackage{amssymb}
\usepackage{amsmath}
\usepackage{amsthm}
\usepackage{thm-restate}
\usepackage{hyperref}
\usepackage[ruled,vlined]{algorithm2e}
\usepackage{cleveref}

\addtokomafont{disposition}{\rmfamily}

\theoremstyle{definition}
\newtheorem{definition}{Definition}
\theoremstyle{plain}
\newtheorem{theorem}[definition]{Theorem}

\newtheorem{proposition}[definition]{Proposition}
\theoremstyle{remark}

\newcommand\claimqedhere{\qed}
\newenvironment{claimproof}{\textit{Proof of the claim.}}{\claimqedhere}

\usepackage{enumitem}
\setlist{nosep}

\usepackage{complexity}
\usepackage[ruled,vlined]{algorithm2e}

\newcommand\EXPTIME{\class{EXPTIME}}
\newcommand\APSPACE{\class{APSPACE}}

\newcommand\dom{\ensuremath{\operatorname{dom}}}
\renewcommand\iff{\,\Leftrightarrow\,}

\title{The Inclusion Problem for \\ Forest Languages under Substitutions} 

\newcommand\affiliation[1]{\\{\small{}#1}}
\newcommand\email[1]{\\{\small{}\href{mailto:#1}{{\small{}\texttt{#1}}}}}
\newcommand\orcid[1]{}

\author{Marcial Gaißert
  \affiliation{University of Stuttgart, FMI, Germany}
  \email{marcial.gaissert@gaisseml.de}
  \orcid{https://orcid.org/0000-0003-0566-7632} \and 
  Manfred Kuf\-leitner
  \affiliation{University of Stuttgart, FMI, Germany}
  \email{kufleitner@fmi.uni-stuttgart.de}
  \orcid{https://orcid.org/0000-0003-3869-416X}}
  
\date{}

\begin{document}
\SetEndCharOfAlgoLine{}

\maketitle

\begin{abstract}
  \noindent
  \textbf{Abstract.}\; 
%
We consider algorithms and lower bounds for various problems over forest languages; as input models we allow forest algebras, deterministic forest automata and nondeterministic forest automata. For the equivalence problem, we give an almost-linear-time algorithm for both forest algebras and deterministic forest automata; this is complemented by a polynomial time hardness result. The emptiness problem is complete for polynomial time over each of the three models. Additionally, we consider the emptiness of intersection problem for forest algebras and deterministic forest automata; this problem turns out to be complete for exponential time. It is well-known that the corresponding problems for word languages are complete for nondeterministic logarithmic space and for polynomial space, respectively.

Equipped with this toolbox of algorithms and lower bounds, we consider various inclusion problems for regular forest languages under substitutions. The substitutions in this paper replace leaf variables by forest languages. Depending on the direction of the inclusion, the problem for a given substitution is either complete for polynomial time or for exponential time; in particular, the equivalence problem under substitutions is complete for exponential time and, hence, more difficult than the equivalence problem for forest languages without substitutions. If we ask whether there exists a substitution such that a given inclusion holds, then this problem is either complete for \NP\xspace or exponential time, depending on whether we consider inclusion or equivalence; moreover, the problem is undecidable if the substitution is applied on both sides.

\end{abstract}

\section{Introduction}
Boja{\'n}czyk and Walukiewicz introduced
the notion of a forest algebra alongside the recognition of forest languages
by finite forest algebras~\cite{bw07lahp}.
They also gave an equivalent but more succinct model called forest automata.
We add to this the naturally occurring model of nondeterministic forest automata.
Forests generalize ordered unranked trees to finite series of unranked trees,
thereby admitting a uniform algebraic structure.

We consider \emph{relational} substitutions on forest languages.
Due to the additional structure of forest languages in contrast to word languages,
multiple variants arise. We restrict ourselves to substituting leaves
independently of one another.

In \cref{sec:constructions} we briefly review some of the usual constructions
on automata or algebraic models with their runtime or space bounds,
and constructions for the aforementioned substitutions.
In \cref{sec:algorithms}, we give an almost linear time algorithm
for the equivalence of deterministic forest automata\,---\,generalizing the Hopcroft-Karp equivalence test\,---\,and a polynomial-time algorithm for checking emptiness of nondeterministic forest automata.
In \cref{sec:problems_without}, we then give completeness results for
the emptiness, subset and equivalence problems on deterministic forest automata and
forest algebras, as well as for the emptiness of the intersection of forest algebras.\footnote{Missing proofs can be found in the appendix of this submission.}

Conway considered substitutions
which replace variables in a word by words from a given language~\cite{conway1971regular}.
He showed that the set of inclusion maximal substitutions \(\sigma\) satisfying \(\sigma(L)\subseteq R\) is finite.
Furthermore, for regular languages \(L\) and \(R\), all maximal substitutions are regular, that is,
the languages by which the variables are substituted
are all regular.
The same result can easily be obtained when describing the languages
using recognizing monoids and applying a method called saturation;
this works by simply extending the substitutions to contain complete classes
of the syntactic monoid of \(R\).

It is natural to ask whether the principle of saturation extends from
word languages recognized by finite monoids
to forest languages recognized by finite forest algebras.
For substitutions of leaves this is the case, as we will show in~\cref{sec:relational}. In addition, we provide matching lower bounds.
Our results hold for both finite forest algebras
and deterministic forest automata.
Additionally, some results extend to nondeterministic forest automata.

There exists several related work in the literature on trees.
For instance, the regular matching problem and the inclusion problem discussed
by Boneva, Niehren and Sakho~\cite{bns2019regmatch}
are similar in spirit to the ones described here.
In~\cite{bns2019regmatch}, one problem is that string patterns can only be embedded vertically,
thus making context variables necessary.
With forests, this problem does not occur,
since we can embed string patterns horizontally.
Delignat-Lavaud and Straubing~\cite{DelignatLavaudStraubing2010} considered constructions
on the more succinct automaton models of BUDFAs and BUNFAs from a more practical perspective.

\section{Preliminaries}
\subsection{Forest Languages, Forest Algebras and Forest Automata}

The following definitions of forests, forest languages, forest automata
and recognition are due to Boja{\'n}czyk and Walukiewicz~\cite{bw07lahp}.
We define the set \(\mathcal{F}(A)\) of forests
and the set \(\mathcal{T}(A)\) of trees
over a finite, non-empty alphabet \(A\)
inductively as follows:
\begin{itemize}
\item If \(f_1,\dots, f_k\in\mathcal{T}(A)\) for \(k \ge 0\),
  then \(f_1+\cdots+f_k\in\mathcal{F}(A)\).

  For \(k=0\) we denote the resulting forest by \(0\)
  and call it the \emph{empty forest}.
\item If \(f\in\mathcal{F}(A)\) and \(a\in A\),
  then \(af\in\mathcal{T}(A)\).
\end{itemize}
For forests \(f=f_1+\cdots+f_k\)
and \(g=g_1+\cdots+g_\ell\)
with \(f_i,g_i\in\mathcal{T}(A)\),
let \(f+g=f_1+\cdots+f_k+g_1+\cdots+g_\ell\).
Furthermore, \(a0=a\).
Note that \(+\) is not required to be commutative.
A forest induces a partial map
\(f\colon\dom(f)\to A\) where \(\dom(f)\) is defined by
\begin{align*}
  \dom(0)
  &= \emptyset \\
  \dom(f_1+\cdots+f_k)
  &= \left\{ ix \,\middle|\, 1x\in\dom(f_i), 1\le i\le k \right\}
  &&\text{for }f_1,\dots,f_k\in \mathcal{T}(A) \\
  \dom(af)
  &= \left\{ 1x \,\middle|\, x\in\dom(f) \right\}
  &&\text{for }a\in A,f\in\mathcal{F}(A)
\end{align*}
The elements of \(\dom(f)\) correspond to positions or nodes in the forest; their images correspond to their label.

A \emph{context} over an alphabet \(A\) is a forest over \(A\mathbin{\dot\cup}\{1\}\)
such that~\(1\) occurs only once, and this occurrence is a leaf.
The set of all contexts is denoted by \(\mathcal{C}(A)\).
For a context \(c\in\mathcal{C}(A)\)
and a forest \(f\in\mathcal{F}(A)\),
we define \(cf\)
as the forest
obtained by replacing
the occurrence of \(1\) in \(c\)
by the forest \(f\).

A forest \(g\in\mathcal{F}(A)\) is a \emph{subforest} of a forest \(f\in \mathcal{F}(A)\)
if there exists a \emph{context} \(c\in\mathcal{C}(A)\) such that \(f = cg\). A \emph{subtree} is a subforest which is a tree,  i.e., a subforest of the form \(af\) for some \(a\in A, f\in \mathcal{F}(A)\).

A forest algebra \((H,V,\cdot,\mathrm{in}_\ell,\mathrm{in}_r)\)
is a 5-tuple consisting of:
\begin{itemize}
\item A \emph{horizontal monoid} \((H,+,0)\)
\item A \emph{vertical monoid} \((V,\cdot,1)\)
\item A monoid action \(\cdot\colon V\times H\to H\) of \(V\) on \(H\).
\item Two functions \(\mathrm{in}_\ell,\mathrm{in}_r\colon H\to V\)
  such that \(\mathrm{in}_\ell(g)h=g+h\)
  and \(\mathrm{in}_r(g)h=h+g\).
\end{itemize}
We also write forest algebras as pairs \((H,V)\), and \(\mathrm{in}_\ell\) and \(\mathrm{in}_r\) are written as \(\mathrm{in}_\ell(g)=g+1\)
and \(\mathrm{in}_r(g)=1+g\). A forest algebra \((H,V)\) is \emph{finite} if both \(H\) and \(V\) are finite.

Over an alphabet \(A\),
the \emph{free forest algebra}
\(A^\Delta\)
is
\((
(\mathcal{F}(A),+,0),
(\mathcal{C}(A),\cdot,1),
\cdot,
\mathrm{in}_\ell,
\mathrm{in}_r
)\)
with \(\mathrm{in}_\ell\) and \(\mathrm{in}_r\) uniquely determined by the rest.
For a forest algebra \((H,V)\) and \(h\subseteq H\)
we let \({h}^* = \left\{ f_1+\cdots+f_k \,\middle|\, k\in\mathbb{N}, f_i\in h \right\}\).
Given two forest algebras \((H,V)\)
and \((G,W)\),
a forest algebra homomorphism
\(\varphi\colon (H,V) \to (G,W)\)
is a pair \((\alpha,\beta)\) of monoid homomorphisms
with \(\alpha(vh)=\beta(v)\alpha(h)\),
\(\beta(\mathrm{in}_\ell(h))=\mathrm{in}_\ell(\alpha(h))\) and
\(\beta(\mathrm{in_r}(h))=\mathrm{in}_r(\alpha(h))\)
for all \(v\in V\) and \(h\in H\).
We then also write
\(\varphi(v)=\alpha(v)\)
and \(\varphi(h)=\beta(h)\)
for \(v\in V\) and \(h\in H\).

A \emph{forest language} over the alphabet \(A\) is a subset \(L\subseteq \mathcal{F}(A)\).
It is \emph{recognized} by
a forest algebra \((H,V)\)
if there exists
a homomorphism \(\varphi\colon A^\Delta\to (H,V)\)
and a set \(E\subseteq H\)
such that for all forests \(f\in\mathcal{F}(A)\)
we have \(f\in L\)
if and only if \(\varphi(f)\in E\).
A forest language is \emph{recognizable}
if it is recognized by a finite forest algebra.

Boja{\'n}czyk and Walukiewicz require the monoid action to be faithful~\cite{bw07lahp},
that is, that for all \(u,v\in V\) with \(u\neq v\),
there exists \(h\in H\) such that \(uh\neq vh\).
Delignat-Lavaud and Straubing~{\cite[Remark 3]{DelignatLavaudStraubing2010}} showed that the class of recognizable forest languages is independent of the
additional requirement of the forest algebra to be faithful: for every forest algebra \((H,V)\) there exists a faithful forest algebra \((H,V/{\sim_H})\) such that every forest language \(L\) recognized by \((H,V)\) is also recognized by \((H,V/{\sim_H})\). The definition of the congruence \(\sim_H\) on \(V\) is given by
\(
  u\sim_H v :\iff \forall h\in H\colon uh=vh
\)
for elements \(u,v\in V\). Then, \((H,V/\sim_H)\) is a faithful forest algebra. 
  The natural homomorphism \(\pi:(H,V) \to (H,V/{\sim_H})\) satisfies \(\varphi(L) = \pi(\varphi(L)) \subseteq H\);
  thus, \(L\) is recognized by \((H,V/{\sim_H})\). For a given forest algebra \((H,V)\) and given elements \(u,v\in V\), one can check in deterministic logarithmic space whether \(u \sim_H v\) for \(u,v\in V\).

A (deterministic) \emph{forest automaton} over an alphabet \(A\) is a tuple
\(M=(
(Q,+,0),
A,
\delta,
F
)\)
such that
\begin{itemize}
\item \((Q,+,0)\) is a finite monoid and its elements are the states,
\item \(\delta\colon A\times Q\to Q\) is the transition function and
\item \(F\subseteq Q\) is the set of accepting states.
\end{itemize}
The evaluation \(f^M\) of a forest \(f \in \mathcal{F}(A)\) under the automaton \(M\) is inductively defined by 
\begin{align*}
  0^M
  &= 0 \\
  {(af)}^M
  &= \delta(a,f^M)
  &&\text{for }a\in A, f\in\mathcal{F}(A) \\
  {(f_1+\cdots+f_k)}^M
  &= f_1^M+\cdots+f_k^M
  &&\text{for }f_1,\dots,f_k\in\mathcal{T}(A)
\end{align*}
The language accepted by \(M\) is \(L(M) = \{ f \in \mathcal{F}(A) \mid f^M\in F \}\).
It will also be helpful to define \(\hat\delta\colon \mathcal{C}(A) \times Q\to Q\) with \(\hat\delta(1, q) = q\),
\(\hat\delta(f+c, q) = f^M + \hat\delta(c,q)\),
\(\hat\delta(c+f, q) = \hat\delta(c,q) + f^M\),
\(\hat\delta(ac, q) = \delta(a, \hat\delta(c, q))\)
for all \(a\in A, c\in \mathcal{C}(A), f\in\mathcal{F}(A)\).

As~\cite{bw07lahp} show,
a forest language
is accepted by a forest automaton
if and only if it is recognized by a finite forest algebra.
Furthermore, the conversion
from a forest algebra
to a forest automaton
is obviously possible
using only logarithmic space.

A \emph{nondeterministic} forest automaton \(M=((Q,+,0),A,\delta,F)\)
is defined like a forest automaton,
except with \(\delta\) being a function to a set of states,
and, accordingly, \(0^M = \{0\}\), \({(af)}^M = \bigcup_{q\in f^M} \delta(a,q)\), \({(f_1 + \cdots+ f_k)}^M = \{q_1+\cdots+ q_k \mid \forall 1\le i\le k\colon q_i\in f_i^M\}\) for all \(f,f_1,\dots,f_k\in\mathcal{F}(A), a\in A\) and \(L(M)=\{f\in\mathcal{F}(A) \mid f^M\cap F\neq\emptyset\}\)

We can encode Boolean formula and expressions \(F\)
(with \(\top,\bot\) already substituted in for the variables,
or with variables from a set \(X\))
as a forest \(\langle F \rangle_{\mathcal{F}}\)
over \(\{\land,\lor,\neg, \bot, \top\}\cup X\) using:
\begin{align*}
  \langle L\land R \rangle_{\mathcal{F}}
  &= \land (\langle L \rangle_{\mathcal{F}} + \langle R \rangle_{\mathcal{F}})
  &&\forall L,R\\
  \langle L\lor R \rangle_{\mathcal{F}}
  &= \lor (\langle L \rangle_{\mathcal{F}} + \langle R \rangle_{\mathcal{F}}) \\
  \langle \neg L \rangle_{\mathcal{F}}
  &= \neg (\langle L \rangle_{\mathcal{F}}) \\
  \langle x \rangle_{\mathcal{F}} &= x &&\forall x\in \{\bot,\top\}\cup X
\end{align*}
Here \(\top\) and \(\bot\) are meant as representations of true and false, respectively.

Now, let
\(\mathrm{True}_{\mathcal{F}}
=\left\{
  \langle F \rangle_{\mathcal{F}}
  \,\middle|\,
  F\text{ is a true Boolean expression without variables}
\right\}\),
which is a recognizable forest language,
using a forest algebra whose horizontal elements
correspond to pairs of boolean values,
with the neutral element being \((\bot,\top)\)
and the horizontal operation combining
the first component using logical disjunction
and the second using conjunction.

\subsection{Substitutions on Forest Languages}
For forests, we define a model of substitutions at the leaves:
\begin{definition}
  Let \(\sigma\colon X\to 2^{\mathcal{F}(A)}\setminus\{\emptyset\}\) with \(X\cap A=\emptyset\).
  Then we can extend \(\sigma\) to \(\sigma_r\) by 
  \[
    \sigma_r\colon \begin{aligned}[t]
      \mathcal{F}(A\cup X) &\;\to\; 2^{\mathcal{F}(A)}
      \\ x &\;\mapsto\; \sigma(x)
      \\ af &\;\mapsto\; \left\{ af' \,\middle|\, f'\in\sigma_r(f) \right\}
      \\ g_1+\cdots+g_k
      &\;\mapsto\;\left\{ g_1'+\cdots+g_k' \,\middle|\, \forall 1\le i\le k\colon g_i'\in\sigma_r(g_i) \right\}
    \end{aligned}
  \]
  for all $x \in X$, $a \in A$, $k \in \mathbb{N}$ and all $f,g_1,\ldots, g_k \in \mathcal{F}(A)$.
  We call the function \(\sigma_r\) a \emph{relational substitution} (at the leaves) or sometimes just a \emph{substitution}.
  Note that it is a partial function insofar that the value for forests containing variables
  (i.e., elements of \(X\)) at inner nodes is not defined.
  Those are also the only forests with undefined values.
\end{definition}
We call a substitution \(\sigma\) \emph{homomorphic} if for all \(x\in X\),
we have \(\left| \sigma(x) \right| = 1\)
and we write \(\sigma\subseteq\theta\) for two substitutions \(\sigma\) and \(\theta\) if,
for all \(x\in X\), we have \(\sigma(x)\subseteq\theta(x)\).
For forest languages \(L\subseteq \mathcal{F}(A)\)
and substitutions \(\sigma\colon\mathcal{F}(A)\to 2^{\mathcal{F}(A)}\),
we write \(\sigma(L) = \bigcup_{f\in L}\sigma(f)\).
Additionally, let \(\sigma_r^{-1}(L)=\{ f \in\mathcal{F}(A) \mid \sigma_r(f) \cap L \neq\emptyset\} = \bigcup_{l\in L}\{f\in\mathcal{F}(A)\mid l\in\sigma_r(f)\}\).

When relational substitutions are part of the input of an algorithm,
we assume the values to be encoded as forest automata or recognizing forest algebras.
Note that this especially means that, for problems with given substitutions,
we only consider substitutions which map variables to recognizable forest languages.

\subsection{Complexity Theory Basics}
In this paper, we show completeness results for problems
in the complexity classes \P, \NP\ and \EXPTIME.\@
For \EXPTIME-completeness,
we use a characterization in terms of alternating turing machines: \(\APSPACE=\EXPTIME\)~\cite{ChandraKozenStockmeyer81}.
For definitions of these classes and an introduction
into complexity theory, the reader may refer to~\cite{AroBar09}.
\section{Basic Constructions}\label{sec:constructions}
First, using the obvious product and complement forest algebra respectively automaton constructions,
we get the following result.
Here, as in the word case, the complement construction requires a determinized automaton,
but constructions for monotone boolean operations can also be efficiently carried out
on nondeterministic forest automata.
\begin{restatable}{proposition}{effectiveBooleanAlgebra}\label{thm:BooleanAlgebra}
  The languages recognizable using forest algebras form an effective Boolean algebra.
  The constructions are possible using only logarithmic space.

  This result also holds for deterministic forest automata.
\end{restatable}
Another useful standard closure property is the closure under inverse homomorphisms:
\begin{restatable}{proposition}{inverseHom}\label{thm:inverseHom}
  Let \(L\in\mathcal{F}(B)\) be a forest language recognized by a forest algebra resp.\ forest automaton
  and \(\varphi\colon A^\Delta \to B^\Delta\) be a forest algebra homomorphism.
  Then we can construct a forest algebra resp.\ forest automaton for \(\varphi^{-1}(L)\)
  in logarithmic space.
\end{restatable}
Also, a nondeterministic forest automaton might be determinized,
constructing a deterministic forest automaton:
\begin{restatable}{theorem}{powersetDFA}\label{thm:powersetDFA}
  Given a nondeterministic forest automaton,
  we can construct
  a deterministic forest automaton of exponential size
  accepting the same forest language.
  This construction is possible in a time polynomial in the size of the output.
\end{restatable}
In contrast to the case for finite word automata, the construction is not possible
using logarithmic space in the output unless \(\PSPACE=\EXPTIME\).
This follows from~\cref{thm:faIntEmp} using a similar construction to the
\PSPACE-hardness of universality for nondeterministic finite word automata.

\subsection{Substitutions}
We will now consider constructions for forest languages gotten using substitutions,
ways to combine inequalities with substitutions and a result to represent unions
succinctly using a substitution.
\begin{theorem}\label{thm:oiNFA}
  Given a forest automaton \(M=((Q,+,0),\delta,E)\)
  and a recognizable relational substitution \(\sigma\),
  we can construct, in polynomial time, a nondeterministic forest automaton \(M'=((Q',+,0'),\delta',E')\)
  such that \(\sigma(L(M)) = L(M')\),
  and \(\left| Q' \right|\) is polynomial.
\end{theorem}
\begin{proof}
  Let \(M=((Q,+,0),\delta,E)\) be a nondeterministic forest automaton
  and \(\sigma\colon X\to 2^{\mathcal{F}(A)}\) a relational substitution
  with \(M_{\sigma(x)}=((Q_{\sigma(x)},+,0_{\sigma(x)}),\delta_{\sigma(x)},E_{\sigma(x)})\)
  being a nondeterministic forest automaton for every \(x\in X\).
  Without loss of generality, assume that for all forests \(f\neq 0\),
  \(0\notin f^M\) and \(0_{\sigma(x)}\notin f^{M_{\sigma(x)}}\) for all \(x\in X\).
  Then we construct the nondeterministic forest automaton \(M'=((Q',+,0'),\delta',E')\)
  as follows.

  The idea of this construction is to let the forest automaton
  guess nondeterministically for each
  leaf, whether this is part of a substituted subtree or a direct child,
  and, at each node in a substituted part, if this is the right-hand side
  corresponding to a \(x\in X\) and thus
  should be replaced by that \(x\) (which is marked by a \(s_x\)).
  
  We describe the state set \(Q'\)
  as a quotient of the free product (in additive notation) of \(S={\{s_x \mid x\in X\}}^*\), \(\{0,\bot\}\) (with \(\bot+\bot=\bot\)), \(Q\) and all sets \(Q_{\sigma(x)}\) for \(x \in X\)
  by:
  \begin{itemize}
  \item \(s_x+e+p = x^M + p\) if \(e\in E_{\sigma(x)}, p\notin Q_{\sigma(x)}\),
  \item \(s_x+b+p = \bot\) if \(b\in Q_{\sigma(x)}\setminus E_{\sigma(x)}, p\notin Q_{\sigma(x)}\),
  \item \(p+q = \bot\) if \(q\in Q_{\sigma(x)}, p\in \left( Q\cup \bigcup_{y\in X} Q_{\sigma(y)} \right) \setminus Q_{\sigma(x)}\) and
  \item \(p+\bot =\bot, \bot+p = \bot \) for all \(p\in Q'\).
  \end{itemize}
  It is important that \(0\) is contained in \(Q\) and \(Q_{\sigma(x)}\) for all \(x\in X\).
  Also, reading those equalities as reduction rules from left to right results in a convergent rewriting system.
  The set of normal forms has polynomial size since all normal forms have length at most \(2\):
  \begin{itemize}
  \item \(\bot\), \(q\) with \(q\in Q\cup\bigcup_{x\in X} Q_{\sigma(x)}\),
  \item \(s_x + q\) with \(q\in Q_{\sigma(x)}\),
  \item \(p + q\) with \(p\in \bigcup_{x\in X} Q_{\sigma(x)}, q\in Q\)
  \end{itemize}

  To simplify the presentation, we define \(r\colon Q' \to Q\cup\{\bot\}\)
  with \(r(q)=q\) for \(q\in Q\), \(r(s_x+p) = x^M\) for \(p\in E_{\sigma(x)}\)
  and \(r(q')=\bot\) for all other \(q'\in Q'\).
  Note that \(r(x+y)=r(x)+r(y)\) if \(y\) is not of the form \(p+q\) with \(p\in Q_{\sigma(x)}, q\in Q, x\in X\).

  Then, we define \(\delta'(q,a) = S + D + S\) with \(D\) containing all elements of:
  \begin{itemize}
  \item \(\delta(q',a)\) if \(q'\in r(S + q + S)\setminus \{\bot\}\) and
  \item \(\delta_{\sigma(x)}(q,a)\) if \(q\in Q_{\sigma(x)}\).
  \end{itemize}
  The accepting states are all \(q\in Q'\) with \(r(S+q+S)\cap E\neq\emptyset\).
  Note that \(f^{M'} = S+f^{M'}+S\) for all \(f\in\mathcal{F}(A)\setminus\{0\}\).

  Using a somewhat technical induction we can now show:
  \begin{restatable}{claim}{oiNFAClaim}
    Let \(f\in\mathcal{F}(A)\) and \(g\in\mathcal{F}(A\cup X)\).
    Then, \(f\in\sigma(g)\) if, and only if, \(g^M\subseteq r(S+f^{M'}+S)\).
  \end{restatable}

  It is easy to see that the \(Q'\), the operation in \(Q'\),
  the set of accepting states,
  and all values of \(\delta'\) can be computed in polynomial time
  when represented using normal forms,
  which then also holds for the nondeterministic forest automaton.
\end{proof}

We can also prove a similar result for the inverse application of substitutions,
which we state in the following.
The significantly simpler proof, nondeterministically guessing the substitution
in the transition for an \(x\in X\), is omitted for brevity.
\begin{restatable}{theorem}{oiInvNFA}\label{thm:oiInvNFA}
  Given a forest automaton \(M=((Q,+,0),\delta,E)\)
  and a recognizable relational substitution \(\sigma\),
  we can construct
  a nondeterministic forest automaton \(M'=((Q,+,0),\delta',E)\) of equal size
  such that \(L(M') = \sigma^{-1}(L(M))\).
\end{restatable}

Often, it can be useful to consider systems of inequalities instead of single
inequalities. This, however, is equivalent for our purposes by the following result,
gotten by distinguishing the single inequalities using an added root node.
\begin{restatable}{proposition}{lemCombineInequalities}\label{lem:combineInequalities}
  Given multiple inequalities of the form \(\sigma(L_i)\subseteq \sigma(R_i)\)
  for \(i\in I\) and some index set \(I\),
  we can construct a single inequality \(\sigma(L)\subseteq \sigma(R)\)
  that is satisfied by those \(\sigma\)
  that satisfy all \(\sigma(L_i)\subseteq \sigma(R_i)\).
  Forest automata or algebras for \(L\) resp.\ \(R\) can be constructed in
  time \(\mathcal{O}\left( \prod_{i\in I} \left| L_i \right| \right)\)
  resp.\ \(\mathcal{O}\left( \prod_{i\in I} \left| R_i \right| \right)\).
\end{restatable}

For the hardness results shown later, we will need to represent a union of forest languages
succinctly using a substitution. This can be achieved by adding marks to all subforests,
distinguishing the languages to be recognized. Those can then be removed using a substitution.
This allows to reduce the required state set from the product to essentially the disjoint union
of the individual state sets, so we obtain:
\begin{restatable}{proposition}{unionAsSubst}\label{thm:unionAsSubst}
  Given recognizable forest languages \(L_1,\dots, L_k\subseteq \mathcal{F}(A)\) using forest algebras resp.\ forest automata,
  we can construct a polynomially-sized homomorphic relational substitution \(\sigma\)
  and a recognizable forest language \(L\) using a forest algebra resp.\ automata
  such that \(L_1\cup \cdots\cup L_k = \sigma(L)\).
\end{restatable}

\subsection{Globally}
For some forest languages, the following construction makes the description simpler:
\begin{definition}\label{def:globalize}
  For a forest language \(L\subseteq \mathcal{F}(A)\),
  the forest languages \(G(L)\) contains all forests \(f\in L\)
  such that
  all subtrees \(ag\) of \(f\)
  with \(a\in A\) and \(g\in\mathcal{F}(A)\) satisfy
  \(g\in L\).
\end{definition}
The \(G\) is a shorthand for ``globally''.
Note that for \(0\notin L\), we immediately have \(G(L)=\emptyset\).
\begin{restatable}{proposition}{globalizeFA}\label{thm:globalize}
  If \(L\) is a recognizable forest language, then so is \(G(L)\).
  Furthermore, the recognizing forest automaton is polynomial in size.
\end{restatable}
The automaton for \(G(L)\) can be constructed by simply checking that
the state of the subforest is accepting for each application of \(\delta\),
and transitioning into a failure state otherwise.

\section{Algorithms}\label{sec:algorithms}
We will now give efficient algorithms for the equivalence of (deterministic) forest automata
and the emptiness of (possibly nondeterministic) forest automata.
\subsection{Equivalence of Deterministic Forest Automata}
The equivalence problem for deterministic forest automata can be decided using
\cref{alg:equivalence}. If the inputs are not equivalent, the algorithm computes a witness.
\begin{algorithm}
  \SetAlgoLined
  \(\mathcal{L} \gets \{(0_1, 0_2, 0)\}\)\;
  \(\mathcal{M} \gets \emptyset\)\;
  \While{\(\mathcal{L}\neq\emptyset\)}{
    Choose \((p_1,p_2,w)\in\mathcal{L}\) and remove this triple from \(\mathcal{L}\)\;
    \If{\(\mathrm{Find}(p_1) \neq \mathrm{Find}(p_2)\)}{
      \eIf{\((p_1,p_2)\in (F\times Q'\setminus F)\cup (Q\setminus F\times F')\)}{
        \Return \(w\)\;
      }{
        \(\mathrm{Union}(p_1,p_2)\)\;
        Add \((p_1,p_2,w)\) to \(\mathcal{M}\)\;
        \For{\(a\in A\)}{
          Add \((\delta_1(p_1,a), \delta_2(p_2,a), aw)\) to \(\mathcal{L}\)\;
        }
        \For{\((q_1,q_2,u)\in \mathcal{M}\)}{
          Add \((q_1+p_1,q_2+p_2,u+w)\) and \((p_1+q_1,p_2+q_2,w+u)\) to \(\mathcal{L}\)\;
        }
      }
    }
  }
  \Return ``\(L(M_1) = L(M_2)\)''\;
  \caption{Equivalence test for forest automata}\label{alg:equivalence}
\end{algorithm}

Let \(\mathrm{Find}_i(p)\) be the result of \(\mathrm{Find}(p)\)
after the \(i\)-th iteration of the outer loop,
and \(\mathrm{Find}(p)\) the result of \(\mathrm{Find}(p)\)
after termination.
To show the correctness of the algorithm we start with two claims about the
generated partition.
\begin{restatable}{claim}{equivClaimOne}\label{claim:1}
  If \(\mathrm{Find}(p_1)=\mathrm{Find}(p_2)\), \(\mathrm{Find}(q_1)=\mathrm{Find}(q_2)\)
  for \(p_1,q_1\in Q_1, p_2,q_2\in Q_2\), then
  \begin{enumerate}
  \item \(\mathrm{Find}(\delta_1(p_1,a))=\mathrm{Find}(\delta(p_2,a))\) for all \(a\in A\).
  \item \(\mathrm{Find}(p_1 + q_1) = \mathrm{Find}(p_2 + q_2)\).
  \end{enumerate}
\end{restatable}
The proof of \cref{claim:1} largely proceeds like for Hopcroft-Karp in the word case,
however, for the second part, we have to argue along a path of added elements.
Let \(L(p) = \left\{ c\in\mathcal{C}(A) \,\middle|\, \hat\delta(p, c)\in E \right\}\).
An induction over the contexts in \(L(p_1)\) resp.\ \(L(p_2)\) yields:
\begin{restatable}{claim}{equivClaimTwo}\label{claim:2}
  If \(\mathrm{Find}(p_1)=\mathrm{Find}(p_2)\), then \(L(p_1)=L(p_2)\).
\end{restatable}

\medskip

This leads us to the actual result:
\begin{theorem}
  If \cref{alg:equivalence} returns ``\(L(M_1)=L(M_2)\)'', then the
  forest automata \(M_1\) and \(M_2\) are equivalent.
\end{theorem}
\begin{proof}
  Using \cref{claim:2} and \(\mathrm{Find}(0_1) = \mathrm{Find}(0_2)\), we obtain \(L(M_1)=L(M_2)\).
\end{proof}
\begin{restatable}{theorem}{equivalenceRuntime}
  \Cref{alg:equivalence} executes at most
  \begin{itemize}
  \item \((m+n-1)\) \(\mathrm{Union}\) operations, and
  \item \(1 + (m+n-1)\cdot(|A| + m+n)\) \(\mathrm{Find}\) operations.
  \end{itemize}
\end{restatable}
Here, the first bound is obvious from the number of individual elements managed by the Union-Find-data structure and
the second bound follows from counting the number of elements added to \(\mathcal{M}\)
for each call to \(\mathrm{Union}\) and evaluating the resulting arithmetic series.
Note that, due to the table for the horizontal operation, the size of the automaton is quadratic
in the number of states. Thus, using an appropriate Union-Find-data structure, we achieve an almost-linear runtime.

\subsection{Emptiness for Nondeterministic Forest Automata}
Emptiness of the language recognized by a nondeterministic forest automaton
can be decided using the simple marking algorithm~\ref{alg:NFAEmp},
which determines the reachable states.
Note that, since \(\mathcal{M}\) only gets bigger, we can reject an input as soon as an
element of \(E\) gets added to \(\mathcal{M}\). This was omitted in the above algorithm for
brevity, and since it does not improve our runtime bound.
\begin{algorithm}
  \SetAlgoLined
  \(\mathcal{M} \gets \{0\}\)\;
  \While{\(\mathcal{M}\) changes}{
    \For{\(q\in\mathcal{M}\)}{
      \For{\(p\in\mathcal{M}\)}{
        \(\mathcal{M} \gets \mathcal{M}\cup \{p+q, q+p\}\)\;
      }
      \For{\(a\in A\)}{
        \(\mathcal{M} \gets \mathcal{M}\cup \delta(q,a)\)\;
      }
    }
  }
  Accept if, and only if, \(\mathcal{M}\cap E=\emptyset\)\;
  \caption{Testing emptiness for a nondeterministic forest automaton}\label{alg:NFAEmp}
\end{algorithm}
\begin{restatable}{proposition}{correctnessNFAEmptiness}
  \Cref{alg:NFAEmp} accepts
  on the input of a nondeterministic forest automaton \(M=\left( (Q,+,0), A, \delta, E \right)\)
  if, and only if,
  \(L(M)=\emptyset\).
\end{restatable}

This algorithm runs at most \(\left| H \right|\cdot\left( 2\left| H \right| + \left| A \right| \right)\)
operations of adding a single element to a set,
since the body of the outer for-loop gets executed at most once for each \(h\in H\),
and executes at most \(2\left| H \right| + \left| A \right|\) such operations.
Also, since any deterministic forest automaton can be easily converted
to a nondeterministic forest automaton, and the problem
for deterministic forest automata is \P-complete by~\cref{thm:withoutSubstitutions},
it is unlikely that a sub-polynomial algorithm exists.

\section{Problems Without Substitutions}\label{sec:problems_without}
Before looking at problems using substitutions,
it is instructive to look at the problems without any substitution.
We will then later use those results directly in reductions
or use similar arguments when considering the same problems
with an added substitution.
\begin{theorem}\label{thm:withoutSubstitutions}
  Given two recognizable forest languages \(L\) and \(R\),
  encoded as forest automata,
  it is \P-complete to decide:
  \begin{enumerate}
  \item\label{itm:withoutSubstitution:empty} \(L=\emptyset\)
  \item\label{itm:withoutSubstitution:subset} \(L\subseteq R\)
  \item\label{itm:withoutSubstitution:equal} \(L=R\)
  \end{enumerate}
  This result also holds if \(L\) and \(R\) are encoded using recognizing finite forest algebras.
\end{theorem}
\begin{proof}
  First, we observe that using \Cref{thm:BooleanAlgebra}, we can reduce these problems
  onto each other as follows:
  \textit{\ref{itm:withoutSubstitution:empty}}\,\(\Rightarrow\)\,\textit{\ref{itm:withoutSubstitution:equal}}\,:
  Let \(R=\emptyset\).
  \textit{\ref{itm:withoutSubstitution:equal}}\,\(\Rightarrow\)\,\textit{\ref{itm:withoutSubstitution:subset}}\,:
  Let \(L'=L\cup R\) and \(R'=L\cap R\).
  We then have
  \( L'\subseteq R' \mathrel{\Leftrightarrow} L\cup R\subseteq L\cap R \mathrel{\Leftrightarrow} L=R\).
  \textit{\ref{itm:withoutSubstitution:subset}}\,\(\Rightarrow\)\,\textit{\ref{itm:withoutSubstitution:empty}}\,:
  by choosing \(L'=L\setminus R\).
  Then
  \( L'=\emptyset \mathrel{\Leftrightarrow} L\setminus R=\emptyset \mathrel{\Leftrightarrow} L\subseteq R \).

  Problem~\textit{\ref{itm:withoutSubstitution:equal}} can be solved in polynomial time using
    \cref{alg:equivalence} given deterministic forest automata.
  Problem~\textit{\ref{itm:withoutSubstitution:empty}} can even be solved in polynomial time
    given nondeterministic forest automata using \cref{alg:NFAEmp}.

  Problem~\textit{\ref{itm:withoutSubstitution:subset}} is \P-hard:
    We proof this by a logspace-reduction from \lang{CircuitValueProblem}.
    The \lang{CircuitValueProblem} asks,
    given a Boolean circuit and inputs \(x_1,\dots,x_k\), whether
    the circuit evaluates to \(\top\)
    and was proven to be \P-complete in~\cite{ladner1975}.
    Note that, to prevent confusion with the empty forest or holes,
    we use \(\bot\) and \(\top\) as values, instead of the often used \(0\) and \(1\).

    Let \(C\) be a Boolean circuit.
    Without loss of generality, we assume that the input nodes
    have already been replaced by their respective values,
    so instead of variables the respective nodes are now \(\top\) or \(\bot\).
    Additionally, we assume all nodes of the boolean circuit to
    be annotated with their position in the input.
    
    This Boolean circuit can now be described as a sequence \((\beta_1,\dots,\beta_n)\)
    where for all \(1\le i\le n\), \(\beta_i\) is one of
    \(\top_i, \bot_i, \land_i(j,k), \lor_i(j,k), \neg_i(j)\)
    with \(j,k<i\).
    
    We then construct a forest algebra \((H,V)\)
    recognizing \(\{{\left\langle F\right\rangle}_{\mathcal{F}}\}\),
    where \(F\) is the Boolean
    formula corresponding to the circuit \(C\),
    with the operators annotated with an index as in the circuit.
    The encoding is defined analogously to the encoding of
    Boolean formulas, but preserving the labels.
    Here, the immediate subformulas
    are always in the order of the position of their
    occurrence in the input or, equivalently,
    their index in the sequence described above.
    We write \(n<n'\) to mean that \(n\) comes before \(n'\) in that order.
    Note that, since the conversion from a recognizing forest algebra
    to a forest automaton is possible using only logarithmic space,
    this shows the result also for forest automata.
    
    While \(F\) may have size exponential in \(C\),
    since the subtrees of gates whose output is used multiple times
    are duplicated, this is not the case for the forest algebra \((H,V)\):

    The elements of horizontal monoid are 
    exactly consecutive sets of nodes in the circuit with a common parent.
    Consecutive for a set \(S\) means that,
    whenever \(n_1,n_3\in S\),
    \(n_2\) is also a child of every common parent of all elements of \(S\),
    and \(n_1<n_2<n_3\),
    then \(n_2\in S\).
    These sets are thus determined
    by the parent and the first and last node in the sequence,
    so there are at most \(n^3\) such elements,
    in addition to the empty set representing a failure state.
    Since we annotated the nodes to have unique labels,
    there cannot be any two such elements that coincide,
    with the exception of equal sets with multiple common parents,
    which we can easily eliminate by checking if there exists
    a previous common parent of the same set of nodes.

    The elements of the vertical monoid are pairs of elements of the
    horizontal monoid \((h_1,h_2)\) with \(h_1\neq h_2\),
    corresponding to functions
    \(H\to H, h_1\mapsto h_2, h\mapsto \emptyset\) for all \(h\in H\setminus\{h_1\}\),
    and additionally, an identity element \(\mathrm{id}_H\).

    The horizontal operation combines \(h_1\) and \(h_2\)
    to \(\emptyset\) except when the elements of \(h_1\) and \(h_2\)
    all have a common parent, are disjoint and are consecutive,
    i.e., \(h_1\cup h_2\) is again an element of the horizontal monoid.
    The vertical operation is function composition.
    
    The forest language \(\mathrm{True}_{\mathcal{F}}'\) of true Boolean formulas
    with arbitrarily annotated operators is recognized by the same forest algebra
    as \(\mathrm{True}_{\mathcal{F}}\),
    by adjusting the homomorphism to ignore the annotations.
    Now, \(\left\{ \langle F \rangle_{\mathcal{F}} \right\}\subseteq \mathrm{True}_{\mathcal{F}}'\) holds
    if and only if the Boolean formula,
    and thus the Boolean circuit,
    evaluates to \(\top\).
\end{proof}

\begin{theorem}\label{thm:faIntEmp}
  Given recognizing finite forest algebras for recognizable forest languages \(L_1,\dots,L_k\)
  it is \EXPTIME-complete to decide whether \(L_1\cap \cdots\cap L_k = \emptyset\).
\end{theorem}
\begin{proof}
  To decide the problem in exponential time, it suffices to construct the
  product automaton of the corresponding forest automata and decide emptiness
  as above.

  To show \EXPTIME-hardness, using
  \(\APSPACE=\EXPTIME\)~\cite{ChandraKozenStockmeyer81},
  we can equivalently show \APSPACE-hardness.
  Let \(L\in\APSPACE\)
  and \(M=(Q,\Gamma,\delta,q_0,g)\) be an alternating turing machine
  with \(L(M)=L\)
  and a polynomial space bound \(p(n)\).
  Without loss of generality, we assume the following restrictions for \(M\):
  \begin{itemize}
  \item In any configuration, there are at most two possible transitions.
  \item \(M\) never moves its head left of the initial position.
  \end{itemize}
  Then, on input of a word \(w=a_1\dots{}a_n\in\Sigma^*\) we construct forest algebras for multiple languages
  over the alphabet \(\Gamma' = \Gamma\times\{1,\dots,p(n)\} \cup \Gamma\times\{1,\dots,p(n)\}\times Q\).
  To make those constructions (and the possibility to construct them efficiently) more readable,
  define the following forest algebra homomorphism from \(\Gamma^\Delta\):
  \begin{align*}
    \mathrm{at}_{I}\colon
    \begin{aligned}[t]
      \Gamma'^\Delta &\to {(\Gamma \cup \Gamma\times Q \cup Q)}^\Delta \\
      \begin{pmatrix}a\\i\end{pmatrix}1 &\mapsto a1 &
      \begin{pmatrix}a\\j\end{pmatrix}1 &\mapsto 1 &&\forall i\in I,j\in\overline{I} \\
      \begin{pmatrix}a\\i\\q\end{pmatrix}1 &\mapsto \begin{pmatrix}a\\q\end{pmatrix}1 &
      \begin{pmatrix}a\\j\\q\end{pmatrix}1 &\mapsto q1
    \end{aligned}
  \end{align*}

  Now, we can define the languages whose intersection is an accepting
  computation tree.
  First, define a language \(\mathrm{Start}\) of all forests whose upmost level is
  \[
    \begin{pmatrix}a_1\\1\\q_0\end{pmatrix}
    \begin{matrix}
    \begin{pmatrix}a_2\\2\end{pmatrix}
    \cdots
    \begin{pmatrix}a_n\\n\end{pmatrix}
    \begin{pmatrix}\square\\n+1\end{pmatrix}
    \cdots
    \begin{pmatrix}\square\\p(n)\end{pmatrix}
    \\\strut
    \end{matrix}
    \text.
  \]

  Then, we define a language \(\mathrm{Accept}\) of all forests where the states contained
  correspond to an accepting computation tree of the alternating turing machine.
  This can be constructed analogous to the construction for \(\mathrm{True}_{\mathcal{F}}\),
  using a homomorphism mapping all \(a1\) where \(a\) does not contain a state to \(1\),
  and the states to \(\land,\lor,\top\) or \(\bot\) according to their type.

  Additionally, we define a language \(\mathrm{Form}\)
  of all forests with the correct syntactic form,
  i.e., all levels are
  \begin{itemize}
  \item in the second component, repetitions of \(1\dots p(n)\)
  \item in the third component, only one state per such repetition.
  \end{itemize}

  Then we construct forest languages for
  checking that we do not modify
  parts of the tape without a state.
  Using the closure under inverse homomorphisms with \(\mathrm{at}_{\{i\}}\) for every \(1\le i\le p(n)\),
  it suffices here to define a single (constant) forest language.

  To check the actual transitions, we again use the closure under inverse homomorphisms
  with \(\mathrm{at}_{\{i-1,i,i+1\}\cap\{1,\dots,p(n)\}}\) for every \(1\le i\le p(n)\).
  This can be applied to a language of the form \(G(\bigcup_{d\in\delta} L_d)\),
  where the languages \(L_d\) checks that the transition \(d\in\delta\) is applied correctly
  from the first to the second level, which is easily seen to be recognizable.
  Note that this also checks that all possible transitions are applied in each state.

  Now, if the intersection of these forest languages is empty, then there is no accepting
  computation tree of \(M\) on input \(w\), and \(w\notin L(M)\).
  On the other hand, if the intersection is non-empty, it contains (at least) one accepting
  computation tree of \(M\) on input \(w\), and \(w\in L(M)\).
\end{proof}

\section{Substitution of Leaves}

\emph{Saturated substitutions} are an important concept for some of the results obtained in the remainder of this section,
whenever we want to decide the existence of a substitution.

\begin{definition}
  Let \(\sigma\) be a relational substitution
  and \(R\) a forest language recognized
  by a forest algebra \((H,V)\) with homomorphism \(\varphi\)
  or accepted by a forest automaton \(M\).
  Then we define the saturated substitutions \(\hat\sigma_{(H,V)}\)
  and \(\hat\sigma_M\) as follows:
  \begin{align*} 
    \hat\sigma_{(H,V)}(x)
    &= \left\{
      f
      \,\middle|\,
      \exists f'\colon f'\in\sigma(x), \varphi(f')=\varphi(f)
    \right\} \\ 
    \hat\sigma_M(x)
    &= \left\{
      f
      \,\middle|\,
      \exists f'\colon f'\in\sigma(x), f'^M=f^M
    \right\}
    \text.
  \end{align*}
  We also write \(\hat\sigma_R\) instead of either of those.
  Note that those notions coincide for a forest algebra
  and the forest automaton constructed from that forest algebra.
\end{definition}

The important property of saturated substitutions is the following:
\begin{restatable}{lemma}{lemSaturation}\label{lem:saturation}
  Let \(\sigma\) be a relational substitution
  and \(L, R\) recognizable forest languages.
  Then, \(\sigma(L)\subseteq R\) if and only if \(\hat\sigma_R(L)\subseteq R\),
  where \(\hat\sigma_R\) is the corresponding saturated substitution.
\end{restatable}
The proof proceeds by induction on the structure of an element of \(L\).
One important corollary from this is that, if there exists a substitution \(\sigma\)
with variables \(X\) such that \(\sigma(L)\subseteq R\), then there also exists a substitution
of polynomial size in the length of the encodings of \(X\) and \(R\) as
forest automata or forest algebras, namely a saturated substitution.

\subsection{Relational Substitutions}\label{sec:relational}
The following theorem forms the basis for all of the complexity upper bounds
shown for problems with relational substitutions.
As noted earlier, since the problems given here get a substitution as an input,
this result only applies to substitutions which restrict the value of a variable
to recognizable forest languages.
\begin{restatable}{theorem}{givenOI}\label{thm:givenOI}
  Given two recognizable forest languages \(L\) and \(R\),
  encoded as forest automata,
  and a recognizable relational substitution \(\sigma\), also encoded using forest automata:
  \begin{enumerate}
  \item\label{itm:givenOI:subset} It is \P-complete to decide if \(\sigma(L)\subseteq R\).
  \item\label{itm:givenOI:supset} It is \EXPTIME-complete to decide if \(\sigma(L)\supseteq R\).
  \item\label{itm:givenOI:equiv} It is \EXPTIME-complete to decide if \(\sigma(L)=R\).
  \item\label{itm:givenOI:subsetboth} It is \EXPTIME-complete to decide if \(\sigma(L)\subseteq \sigma(R)\).
  \end{enumerate}
  The same result holds if the inputs are encoded using finite forest algebras.
\end{restatable}
The problems can be solved using~\cref{thm:withoutSubstitutions}, \cref{alg:NFAEmp} and~\cref{thm:oiNFA},
with the hardness results following from simple reductions from the problems in~\cref{thm:withoutSubstitutions}
respectively~\cref{thm:faIntEmp}, in the latter case using~\cref{thm:unionAsSubst}.

This now allows us to show the main result for problems with relational substitutions.
Note that by~\Cref{lem:saturation}, a relational substitution fulfilling the inequalities
with substitutions only on one side
exists if and only if a saturated one exists. This means that the following result holds for
the existence of arbitrary --- not only of recognizable --- relational substitutions.
\begin{restatable}{theorem}{existsOI}\label{thm:existsOI}
  Given two recognizable forest languages \(L\) and \(R\),
  encoded as forest automata:
  \begin{enumerate}
  \item\label{itm:existsOI:subset} It is \NP-complete to decide if there exists
    a substitution \(\sigma\) such that \(\sigma(L)\subseteq R\).
  \item\label{itm:existsOI:equal} It is \EXPTIME-complete to decide if there
    exists a substitution \(\sigma\) such that \(\sigma(L) = R\).
  \item\label{itm:existsOI:equalboth} It is \emph{undecidable} whether there exists
    a substitution \(\sigma\) such that \(\sigma(L) = \sigma(R)\).
  \end{enumerate}
  The same results hold if the inputs are encoded as finite forest algebras.
\end{restatable}
Those problems may be solved using the results from~\cref{thm:givenOI},
using~\cref{lem:saturation} to bound their size.
Hardness results are trivial reductions from the corresponding problems
in~\cref{thm:givenOI}, using~\cref{lem:combineInequalities}.
The result for~\textit{\ref{itm:existsOI:equalboth}} holds already for star-free word languages~\cite{Kunc07},
and systems of equations over unary word languages~\cite{JezOkhotin2016,LehtinenOkhotin2010} or the existence of regular solutions~\cite{Okhotin2005}.
For forest languages, a system of equations over unary languages can be reduced to a single equation over unary forest languages.
\section{Summary and Open Problems}

We show \P-completeness of emptiness, inclusion and equivalence
of deterministic forest automata and forest algebras. Each of these problems is \NL-complete for word languages~\cite[Thm.\ 26]{JonesDfaEmp}. 
For proving membership in \P\xspace over forest automata, we give efficient algorithms; for instance, we generalize the Hopcroft-Karp equivalence test to deterministic forest automata. Another basic problem in automata theory is deciding the emptiness of the intersection; here, we show that the problem is \EXPTIME-complete for forest algebras. The corresponding problem for word languages is \PSPACE-complete~\cite[Lemma\ 3.2.3]{KozenDfaIntEmp}.

We also consider the inclusion and the equivalence problems in the presence of (relational) substitutions. Depending on which side of an inclusion we allow substitutions, the problem is either \P-complete or \EXPTIME-complete. The latter complexity also applies to the equivalence problem.
When asking whether there exists a substitution such that some inclusion or some equivalence holds under this substitution, the inclusion problem is \NP-complete and the equivalence problem is \EXPTIME-complete. If we allow substitutions on both sides, we can encode language equations and, hence, the problem is undecidable.


In future research, it would be interesting to consider the above problems under more expressive models of substitution such as~\cite{bns2019regmatch} where substitutions can also be applied to inner nodes. Another interesting mode of substitution is obtained from the restriction where the substitution is applied more synchronously: identical variables need to be replaced by identical forests from a given language (rather than by identical forest languages).
\bibliography{traces,bibliography}

\newpage
\appendix
\section{Details of the Basic Constructions}
Most of the details and correctness proofs
for the constructions in were omitted in~\cref{sec:constructions}.
In this section, we will give the concrete substitutions and
correctness proofs for the results provided here.

\subsection{Closure under Boolean Operations}
For boolean operations, as in the word case, we consider
constructions for monotonic boolean operations
and complementation separately, with the latter only possible
for deterministic models.
\begin{proposition}\label{thm:productFA}
  Given recognizable forest languages \(L_1\) and \(L_2\),
  the forest languages \(L_1\cup L_2\) and \(L_1\cap L_2\) are
  effectively recognizable, i.e., we can construct
  them in logarithmic space,
  for deterministic and nondeterministic forest automata and forest algebras.
\end{proposition}
\begin{proof}
  For \(i\in\{1,2\}\),
  let \(L_{i}\) be forest languages recognized by
  the forest algebras \((H_{i},V_{i})\) using the homomorphism \(\varphi_{i}\)
  and recognizing set \(E_{i}\).

  Then the product forest algebra (without faithfulness)
  \((H_1\times H_2, V_1\times V_2)\) with all operations defined element-wise
  recognizes \(L_1\cap L_2\) and \(L_1\cup L_2\) using the homomorphism
  \(\varphi_1\times\varphi_2: x \mapsto (\varphi_1(x),\varphi_2(x))\)
  and recognizing set \(E_1\times E_2\)
  resp.\ \(E_1\times H_2\cup H_1\times E_2\).

  Analogous constructions are possible for languages accepted by
  deterministic or nondeterministic forest automata.
\end{proof}
\begin{proposition}\label{thm:complementFA}
  Given a recognizable forest language \(L\),
  the forest language \(\mathcal{F}(A)\setminus L\) is
  effectively recognizable, i.e., we can construct
  it in logarithmic space,
  for deterministic forest automata and forest algebras.
\end{proposition}
\begin{proof}
  Let \(L\) be forest languages recognized by
  the forest algebras \((H,V)\) using the homomorphism \(\varphi\)
  and recognizing set \(E\).

  Then the complement is recognized by the forest algebra \((H_1,V_1)\)
  using \(\varphi\) and recognizing
  set \(H\setminus E\).

  An analogous construction is possible for languages accepted
  by deterministic forest automata.
\end{proof}
As a corollary of those two propositions, we now get the one given in the main text:
\effectiveBooleanAlgebra*
\subsection{Determinization}
Nondeterministic forest automata can be determinized to deterministic forest automata
analogously to the word case:
\powersetDFA*
\begin{proof}
  Let \(M=((Q,+,0),\delta,E)\) be a nondeterministic forest automaton.
  Then we can construct \(M_{\mathrm{det}}=((2^Q,+,\{0\}),\delta_{\mathrm{det}},E_{\mathrm{det}})\) with
  \begin{align*}
    P_1 + P_2 &= \{ p_1+p_2 \mid p_1\in P_1, p_2\in P_2\} &&\text{for }P_1,P_2\in 2^Q\\
    \delta_{\mathrm{det}}(P, a) &= \bigcup_{p\in P} \delta(p, a) \\
    E_{\mathrm{det}} &= \{ P\in 2^Q \mid P\cap E\neq\emptyset \}
  \end{align*}
  The time bound can be achieved by constructing all states reachable from \(\{0\}\)
  using the above operations until there are no more changes.
\end{proof}
\subsection{Closure under Inverse Homomorphisms}
(Effective) closure under inverse homomorphisms is easy to proof:
\inverseHom*
\begin{proof}
  Given a recognizing forest algebra, we can simply compose the two homomorphisms.
  For forest automata, we can compute a new \(\delta'\)
  using \(\delta'(a,x) = \hat\delta(\varphi(a1), x)\).
  Both of those operations can be carried out element-by-element on the respective function tables.
\end{proof}
\subsection{Substitutions}
In the proof of~\cref{thm:oiNFA}, we omitted the somewhat technical proof of correctness, which we will now give here:
\oiNFAClaim*
\begin{claimproof}
  Now, we first show by induction on \(g\) that,
  if \(f\in \sigma(g)\), then \(g^M\subseteq r(S+f^{M'}+S)\).
  For \(g=0\), we immediately have \(g^M = 0^M = \{0\} = \{r(0)\} \subseteq r(S+0^{M'}+S) = r(S+0+S) = r(S)\) since \(0\in S\).
  For \(g=x\in X\) we have \(x^M \subseteq r(s_x + E_{\sigma(x)}) \subseteq r(S + f^{M_{\sigma(x)}} + S) \subseteq r(S + f^{M'} + S)\) since \(M_{\sigma(x)}\) can be embedded into \(Q'\) and \(\delta_{\sigma(x)}(q,a) \in \delta(q,a)\) for all \(q\in Q_{\sigma(x)}, a\in A\).
  For \(g=ag'\) with \(a\in A, g'\in\mathcal{F}(A)\), we have \(f=af'\) with some \(f'\in\sigma(g')\). By induction hypothesis, \({g'}^M\subseteq r(S+{f'}^{M'}+S)\).
  Then, \({(ag')}^M = \delta({g'}^M, a) \subseteq \delta'({f'}^{M'},a) = {(af')}^{M'}\) since \({g'}^{M'}\in r(S + {f'}^{M'} + S)\setminus\{\bot\}\).
  For \(g=g_1+g_2\) with \(g_1,g_2\in \mathcal{F}(A)\setminus\{0\}\),
  we inductively have \(f_i\in\sigma(g_i)\)
  with \(g_i^M \subseteq r(S + f_i^{M'} + S)\) for \(i\in\{1,2\}\).
  Thus, \(g^M = g_1^M + g_2^M \subseteq r(S+f_1^{M'}+S) + r(S+f_2^{M'}+S)
  = r(S + f_1^{M'} + S + f_2^{M'} + S) = r(S + f_1^{M'} + f_2^{M'} + S)\)
  where the last step is easily seen by considering separately
  the cases where \(f_j=0\) for one or both \(j\in\{0,1\}\).

  On the other hand, show that, if \(q\in r(S+f^{M'}+S)\setminus\{\bot\}\),
  there exists a forest \(g\) with \(q\in g^M\)
  and \(f\in \sigma(g)\).
  If \(f=0\), we have \(0\in f^{M'}\), so \(q\in r(S)\)
  and \(q=r(s_{x_1} + \cdots + s_{x_k}) = r(s_{x_1}) + \cdots + r(s_{x_k})
  = x_1^M + \cdots x_k^M = {(x_1 + \cdots + x_k)}^M\)
  for \(x_1,\dots,x_k\in X, k\in\mathbb{N}\)
  with \(0\in E_{\sigma(x_i)}\) for all \(1\le i\le k\),
  so \(0\in \sigma(x_1+\cdots + x_k)\).
  Now assume \(f\neq 0\).
  Then we have \(S+f^{M'}+S = f^{M'}\) and \(q\in r(f^{M'})\setminus\{\bot\}\).
  We can then write \(f=f_1+\cdots f_\ell\) and \(q=r(q_1+\cdots+q_\ell)\)
  for some \(\ell>0\) such that,
  for all \(1\le i\le \ell\),
  \(q_i\in f_i^{M'}\) and \(q_i\) is either
  from \(s_x + E_{\sigma(x)}\) for some \(x\in X\)
  or \(q_i=\delta(q_i',a)\) for some \(a\in A, q_i'\in r(S + {f_i'}^{M'} + S)\).
  In the former case, we have \(f_i\in\sigma(x)\), and choose \(g_i = x\).
  In the latter case, we inductively get a \(g_i'\)
  such that \({(g_i')}^M=q_i'\) and \(f_i' \in \sigma(g_i')\),
  and choose \(g_i = ag_i'\).
  Then, using \(g=g_1+\cdots g_\ell\), we have \(g^M = q\)
  and \(f = f_1+\cdots f_\ell\in \sigma(g_1) + \cdots + \sigma(g_\ell)
  = \sigma(g_1 + \cdots + g_\ell) = \sigma(g)\).
\end{claimproof}
\subsection{Inverse Substitutions}
We will now give the concrete construction for a nondeterministic forest automaton
recognizing the result of inversely applying a relational substitution.
\oiInvNFA*
\begin{proof}
  Let \(M=((Q,+,0),\delta,E)\) be a nondeterministic forest automaton
  and \(\sigma\colon X\to 2^{\mathcal{F}(A)}\) a relational substitution
  with \(M_{\sigma(x)}=((Q_{\sigma(x)},+,0_{\sigma(x)}),\delta_{\sigma(x)},E_{\sigma(x)})\)
  being a nondeterministic forest automaton for every \(x\in X\).
  Then we construct the nondeterministic forest automaton \(M'=((Q,+,0),\delta',E)\)
  with
  \begin{align*}
    \delta' &\colon \begin{aligned}[t]
      Q \times A &\to 2^Q \\
      q, a &\mapsto \delta(q,a) &&\forall q\in Q, a\in A\setminus X\\
      0, x &\mapsto \bigcup\{f^M \mid f^{M_{\sigma(x)}} \in E_{\sigma(x)}\} &&\forall x\in X
    \end{aligned}
  \end{align*}
  This nondeterministically choses the substitute whenever reading a variable.
  By induction over the forest structure, one can easily show that
  \(f^{M'}\) contains an accepting state if, and only if,
  \(M\) accepts an element of \(\sigma_{r}(f)\), i.e., if \(f\in\sigma_r^{-1}(L(M))\).
\end{proof}
\subsection{Globally}
A forest automaton for \(G(L)\) given some recognizable forest language
\(L\) can be constructed as follows:
\globalizeFA*
\begin{proof}
  Given a forest automaton \(M=((Q,+,0),A,\delta,E)\) for \(L\),
  we add a state \(\bot\) with \(\bot+x=x+\bot=\bot\) for all states \(x\) and
  modify \(\delta\) to \(\delta'(q, a) = \delta(q,a)\) if \(q\in E\)
  and \(\delta(q,a)=\bot\) otherwise, resulting in a forest automaton \(M'\).

  Then, if a forest \(f\) is accepted by this automaton,
  then \(f\in L\), since all transitions that do not lead to a result of \(\bot\)
  were also possible in the automaton for \(L\).
  Additionally, since \(f^{M'}\neq\bot\), this holds also for each subtree.
  Thus, for each subtree \(af'\) of \(f\) with \(a\in A, f'\in\mathcal{F}(A)\),
  we have \({(af')}^{M'} = \delta'({f'}^{M'}, a)\neq \bot\),
  and thus \({f'}^{M'}\in E\) and \({f'}\in L\).

  On the other hand, if a forest \(f\) is in \(G(L)\),
  we can show that \(f^{M'}=f^M\in E\).
  First observe that if \(f\in G(L)\), then \(f'\in G(L)\)
  for all subtrees \(af'\) of \(f\) with \(a\in A, f'\in\mathcal{F}(A)\).
  Now the proof proceeds by induction on \(f\).
  Let \(f = a_1f_1 + \cdots + a_{k}f_k\) for forests \(f_1,\dots,f_k\in\mathcal{F}(A)\).
  Then, we have \(f_i\in G(L)\) for each \(1\le i\le k\) and inductively \(f_i^{M'}\in E\),
  so \(f_i^{M'}\).
  Given that \(f^{M'} = \delta'(f_1^{M'},a) + \cdots + \delta'(f_k^{M'}, a)
  = \delta'(f_1^M, a) + \cdots + \delta(f_k^M, a)
  = \delta(f_1^M, a) + \cdots + \delta(f_k^M, a) = f^M \in E\).
\end{proof}
\subsection{Combining Inequalities}
We obtain the result that allows us to combine multiple inequalities using the following construction.
\lemCombineInequalities*
\begin{proof}
  We can construct \(L\) and \(R\)
  as follows:
  \[
    L=\left\{ i\ell_i \,\middle|\, i\in I, \ell_i\in L_i \right\}
    \qquad\quad R=\left\{ ir_i \,\middle|\, i\in I, r_i\in R_i \right\}
  \]
  The construction in the given time bound is easily possible as a simple extension
  of the product automaton or algebra.
\end{proof}
\subsection{Using Substitutions to Represent Unions}
One construction central to the \EXPTIME-hardness results was that
we can use substitutions to represent unions of forest languages succinctly:
\unionAsSubst*
\begin{proof}
  Assume forest automata \(M_i=((Q_i,+_i,0_i),\delta_i,E_i)\) for \(L_i\) with \(1\le i\le k\)
  and the \(Q_i\) pairwise disjoint.
  Then we can construct \(M=((Q, +, 0), \delta, E)\) with
  \begin{align*}
    Q &= \{0,\bot\} \cup \bigcup_{1\le i\le k} Q_i\times \{i, i'\} \\
    E &= \bigcup_{1\le i\le k} E_i\times \{i\} \\
    (p, i) + (q, j) &= (p +_i q, i) &&i\in\{1,\dots, k\}, j\in \{i,i'\} \\
    (p, i') + (q, j) &= (p +_i q, i') &&i\in\{1,\dots, k\}, j\in \{i,i'\} \\
    0 + x = x + 0 &= x &&x\in Q \\
    x + y &=\bot &&\text{unless defined otherwise} \\
    \delta((q,i'), a) &=( \delta_i(q,a), i ) &&i\in \{1,\dots,k\} \\
    \delta(0, m_i) &= (0_i, i')
  \end{align*}
  This expects a marker for the appropriate \(L_i\) as the leftmost child
  of every node in the forest.

  The forest language recognized by this forest automaton is mapped to \(L_1\cup\cdots\cup L_k\)
  using the substitution \(\sigma\colon \{m_1,\dots, m_k\}\mapsto A\) with \(\sigma(m_i)=\{0\}\).

  For the construction on forest algebras, the main observation is that any context that does not
  contain any of \(m_1,\dots, m_k\) does either have the form \(1\) or \(a1\) for some \(a\in A\)
  or corresponds to the constant \(\bot\) function.
\end{proof}

\section{Omitted Proofs for Algorithms}
Earlier, we gave an algorithm for the equivalence of deterministic forest automata
and the emptiness of nondeterministic forest automata.
Here, we will expand on some parts of the correctness and runtime proofs.
\subsection{Equivalence of Deterministic Automata}
For the equivalence test on deterministic forest automata, we will now expand on the details of the correctness proof, and show the stated runtime bound.
In the correctness proof of~\cref{alg:equivalence}, we omitted the proof of the following two claims:
\equivClaimOne*
\begin{claimproof}
  Assume the contrary and choose \(i\) minimal such that
  there exist \(p_1,q_1\in Q_1, p_2,q_2\in Q_2, a\in A\) with
  \(\mathrm{Find}_i(p_1)=\mathrm{Find}_i(q_1), \mathrm{Find}_i(p_2)=\mathrm{Find}_i(q_2)\)
  and
  \begin{enumerate}
  \item \(\mathrm{Find}_i(\delta_1(p_1,a))\neq\mathrm{Find}_i(\delta_2(p_2,a))\).
    Since \(i\) is minimal, \(\mathrm{Find}_{i-1}(p_1)\neq\mathrm{Find}_{i-1}(p_2)\).
    Thus, \(\mathrm{Find}_i(p_1)=\mathrm{Find}_i(p_2)\) is due to a call to \(\mathrm{Union}\)
    in the \(i\)-th iteration.
    Then the triple \((p_1,p_2,u)\) for some \(u\) was added to \(\mathcal{L}\).
    Since we eventually have \(\mathrm{Find}(z_1)=\mathrm{Find}(z_2)\)
    for all \((z_1,z_2,w)\) added to \(\mathrm{L}\),
    this is a contradiction.
  \item \(\mathrm{Find}_i(p_1+q_1)\neq\mathrm{Find}_i(p_2+q_2)\).
    Since \(i\) is minimal, \(\mathrm{Find}_{i-1}(p_1)\neq\mathrm{Find}_{i-1}(p_2)\)
    or \(\mathrm{Find}_{i-1}(q_1)\neq\mathrm{Find}_{i-1}(q_2)\).
    First, assume \(\mathrm{Find}_{i-1}(p_1)\neq\mathrm{Find}_{i-1}(q_1)\).
    Now, there exist \((z_0,z_1,u_1), (z_1,z_2,u_2), \dots, (z_{k-1},z_k,u_k)\in\mathcal{M}\) with \(z_0=q_1\) and \(z_k=q_2\).
    Then, we add triples \((p_1 + z_i, p_1 + z_{i+1}, u + u_i)\) resp.\ \((p_2 + z_i, p_1 + z_{i+1}, u + u_i)\)
    to \(\mathcal{L}\). Thus, we eventually have:
    \(\mathrm{Find}(p_1 + q_1)=\mathrm{Find}(p_1 + z_0)=\mathrm{Find}(p_2+z_1)=\cdots=\mathrm{Find}(p_2+z_k)=\mathrm{Find}(p_2+q_2)\),
    which is a contradiction.
    The case for \(\mathrm{Find}_{i-1}(q_1)\neq\mathrm{Find}_{i-1}(q_2)\) is symmetric. \claimqedhere
  \end{enumerate}
\end{claimproof}
\equivClaimTwo*
\begin{claimproof}
  Show \(c\in L(p_1) \iff c\in L(p_2)\) by induction on the size of \(c\).
  For \(c=1\), the result is immediately obvious.

  Assume \(c=c'(a1)\) for \(c'\in\mathcal{C}(A)\) and \(a\in A\).
  By \cref{claim:1}, we have \(\mathrm{Find}(\delta_1(p_1,a))=\mathrm{Find}(\delta_2(p_2,a))\) and thus,
  by induction hypothesis, \(c'\in L(\delta_1(p_1,a))\iff c'\in L(\delta_2(p_2,a))\).\

  Now let \(c=c'(f+1)\) or \(c=c'(1+f)\) for some \(c'\in\mathcal{C}(A)\) and \(f\in\mathcal{F}(a)\setminus\{0\}\).
  Then, \(\mathrm{Find}(f^{M_1})=\mathrm{Find}(f^{M_2})\) and thus, \(\mathrm{Find}(f^{M_1} + p_1)=\mathrm{Find}(f^{M_2} + p_2)\)
  by claim 1. Using the induction hypothesis and rewriting concludes the proof.
\end{claimproof}

Also, we only stated the runtime bound without proof:
\equivalenceRuntime*
\begin{proof}
  The first bound is obvious because we only unite disjoint subsets.

  For the second bound, we first note that every \(\mathrm{Find}\) is due to some
  triple in \(\mathcal{L}\). After the \(\mathrm{Union}\),
  we call add \(\left| A \right| + 2M\) elements to \(\mathcal{L}\), where
  \(M\) is the current number of elements in \(\mathcal{M}\).
  In total,
  this leads to \(f\) additions to \(\mathcal{L}\) where
  \begin{align*}
    f &= \sum_{i=1}^{\ell} \left| A \right| + 2M_i
      \le \sum_{i=1}^{\ell} \left| A \right| + 2i
      \le \sum_{i=1}^{n+m-1} \left| A \right| + 2i \\
      &= (n+m-1)\cdot\left| A \right| + (n+m-1)\cdot(n+m) \\
      &= (n+m-1)\cdot(\left| A \right| + n + m)
  \end{align*}
  when \(M_i\) is the size of \(\mathcal{M}\) after the \(i\)-th call to \(\mathrm{Union}\)
  and \(\ell\) is the total number of such calls.
  Together with the initial element, we get \(1+f = 1 + (n+m-1)\cdot(\left| A \right| + n + m)\) calls to \(\mathrm{Find}\).
\end{proof}
\subsection{Emptiness for Nondeterministic Forest Automata}
For emptiness of nondeterministic automata, we omitted the correctness proof,
which we will now give here.
\correctnessNFAEmptiness*
\begin{proof}
  If the algorithm does not accept an input, then after its execution,
  there exists some \(e\in \mathcal{M}\cap E\).
  By straightforward induction over the execution of the algorithm,
  it is easily seen that for each \(m\in\mathcal{M}\),
  there exists a forest \(f\in\mathcal{F}(A)\) such that \(m\in f^M\).
  This also holds for \(e\), so there exists \(f\in\mathcal{F}(A)\) with \(f^M\cap E\neq\emptyset\)
  and \(f\in L(M)\).

  On the other hand,
  assume there exists some forest \(f\in\mathcal{F}(A)\)
  and some \(q\in Q\) with \(q\in f^M\).
  By induction, show that \(q\in\mathcal{M}\) after the execution of the algorithm.
  If \(f=0\), this is obvious.
  If \(f=f_1+f_2\) for two smaller forests \(f_1\in\mathcal{F}(A)\) and \(f_2\in\mathcal{F}(A)\),
  then \(q=q_1+q_2\) for some \(q_1\in f_1^M\) and \(q_2\in f_2^M\).
  Inductively, \(q_1,q_2\in\mathcal{M}\). In the first iteration before which both \(q_1\in\mathcal{M}\) and \(q_2\in\mathcal{M}\),
  we then add \(\{q_1+q_2,q_2+q_1\}\) to \(\mathcal{M}\), and thus \(q=q_1+q_2\in\mathcal{M}\).
  If \(f=af'\) for some \(a\in A, f'\in\mathcal{F}(A)\), we have \(q\in\delta(q',a)\) for some \(q'\in f'^{M}\).
  Inductively, \(q'\in\mathcal{M}\). In the first iteration before which \(q'\in\mathcal{M}\),
  we add all elements of \(\delta(q',a)\) to \(\mathcal{M}\), thus also \(q\in\delta(q',a)\).
\end{proof}

\section{Proof of the Saturation Lemma}
In this section, we show the following --- somewhat technical --- lemma for inclusion
and saturated substitutions that allows us to limit the size of substitutions that
fulfill some inequality with a substitution on only one of the sides.
\lemSaturation*
\begin{proof}
  First, we note that, by definition of \(\hat\sigma\), we have \(\sigma(L)\subseteq\hat\sigma(L)\)
  and thus, the direction from right to left is immediately obvious.

  Let \(R\) be recognized by a finite forest algebra \((H,V)\)
  using a homomorphism \(\varphi\) and recognizing set \(E\),
  and let \(\varphi(L)\subseteq R\).
  Now, for every \(x\in X\) and \(r\in\varphi(\sigma(x))\),
  we choose a representative \(s_{x,r}\in \sigma(x)\)
  with \(\varphi(s_{x,r})=r\). This exists by the definition of \(\hat\sigma_R\).

  We now show for every \(f_L\in\mathcal{F}(A\cup X)\)
  with variables only occuring at the leaves
  and every \(\hat{f}_R\in\hat\sigma_R(f_L)\),
  there exists \(f_R\in\mathcal{F}(A)\) such that
  \(f_R\in\sigma(f_L)\) and \(\varphi(\hat{f}_R)=\varphi(f_R)\).
  For \(0\) this is immediately obvious.
  For the case \(f_L=x\in X\), we can choose \(f_R=s_{x,r}\)
  with \(r=\varphi(\hat{f}_R)\).
  Now, if \(f_L=af_L'\) for some \(f_L'\in\mathcal{F}(A\cup X)\)
  with variables only at the leaves,
  we have \(\hat{f}_R=a\hat{f}_R'\) for some \(\hat{f}_R'\in\hat\sigma(f_L')\) by the
  definition of relational substitutions.
  Inductively, there exists \(f_R'\in\sigma(f_L')\) with \(\varphi(\hat{f}_R')=\varphi(f_R')\),
  thus now \(\varphi(\hat{f}_R)=\varphi(a\hat{f}_R') = \varphi(af_R') = \varphi(f_R)\)
  with \(f_R=af_R'\).
  In the case \(f_L=f_{L1}+\cdots+f_{Lk}\)
  with forests \(f_{L1},\dots,f_{Lk}\in\mathcal{F}(A\cup X), k\in\mathbb{N}\),
  there again exist \(\hat{f}_{R1},\dots,\hat{f}_{Rk}\)
  with \(\hat{f}_{Ri}\in\hat\sigma(f_{Li})\) for \(1\le i\le k\)
  and \(\hat{f}_R=\hat{f}_{R1}+\cdots+\hat{f}_{Rk}\).
  Thus, inductively there exist \(f_{Ri}\) with \(\varphi(f_{Ri})=\varphi(\hat{f}_{Ri})\)
  for \(1\le i\le k\)
  and thus with \(f_R=f_{R1}+\cdots+f_{Rk}\),
  we get \(
  \varphi(\hat{f}_R)
  =\varphi(\hat{f}_{R1})+\cdots+\varphi(\hat{f}_{Rk})
  =\varphi(f_{R1})+\cdots+\varphi(f_{Rk})
  =\varphi(f_R)
  \).

  Now, assume there exists \(\hat{f}_R\in\hat\sigma(L)\setminus R\).
  Then, \(\hat{f}_R\in\hat\sigma(f_L)\) for some \(f_L\in L\),
  and thus there exists also \(f_R\in\sigma(f_L)\) with \(\varphi(\hat{f}_R)=\varphi(f_R)\).
  Using this and \(\hat{f}_R\notin R\), we can now conclude \(f_R\in\sigma(L)\setminus R\).
\end{proof}
\section{Proofs for Problems With Relational Substitutions}
Now, we will show how the other results of the paper can be put together to obtain
the completeness results in~\cref{sec:relational}
\givenOI*
\begin{proof}
  We get \P-hardness of~\textit{\ref{itm:givenOI:subset}} by reduction from the corresponding problems in~\Cref{thm:withoutSubstitutions},
  choosing \(X=\emptyset\),
  and solve problem~\textit{\ref{itm:givenOI:subset}} directly using~\cref{thm:oiNFA} and then deciding \(\sigma(L)\cap \overline{R}=\emptyset\)
  using algorithm~\ref{alg:NFAEmp}.

  Problem~\textit{\ref{itm:givenOI:supset}} is \EXPTIME-hard by reduction from~\cref{thm:faIntEmp}, since emptiness of intersection is equivalent to the universality of a union using the closure under complement, and we can denote the union using a substitution by~\cref{thm:unionAsSubst}.
  Since \(\sigma(L)\subseteq R\) holds trivially in this construction, this also shows that~\textit{\ref{itm:givenOI:equiv}} is \EXPTIME-hard.
  Problem~\textit{\ref{itm:givenOI:subsetboth}} immediately follows by reduction from \textit{\ref{itm:givenOI:supset}},
  since \(\sigma(L)=L\) for \(L\subseteq\mathcal{F}(A)\) (that is, not containing any \(x\in X\)).

  To solve~\textit{\ref{itm:givenOI:supset}} or \textit{\ref{itm:givenOI:subsetboth}} in exponential time, we can simply construct a NFA as in~\cref{thm:oiNFA}, do a powerset construction (\cref{thm:powersetDFA}) and then check the subset relation like in~\cref{thm:withoutSubstitutions}.

  Problem~\textit{\ref{itm:givenOI:equiv}} is solvable in exponential time by solving problems~\textit{\ref{itm:givenOI:subset}}
  and~\textit{\ref{itm:givenOI:supset}} as above.
\end{proof}
\existsOI*
\begin{proof}
  To prove that the problems are in \NP\ resp.\ \APSPACE, we can simply guess a saturated substitution
  and check if the respective inequality holds using the respective
  algorithm from~\Cref{thm:givenOI}.
  Note that the saturated substitutions have right hand sides that are recognized by
  forest automata, have polynomial size in \(L\) and \(R\),
  and every substitution satisfying the inequality can be converted to a saturated one
  using~\Cref{lem:saturation}.
  This also holds for problem~\textit{\ref{itm:existsOI:equal}} since \(\sigma(L)=R\)
  and \(\sigma(L)\subseteq\hat\sigma(L)\subseteq R\) implies \(\hat\sigma(L)=R\).

  We first show that problem~\textit{\ref{itm:existsOI:subset}} is \NP-hard
  by reduction from \lang{Satisfiability}.
  Given a Boolean formula \(F\), we construct its encoding as a forest \({\left\langle F\right\rangle}_{\mathcal{F}}\)
  and a recognizing forest algebra for \(L=\{{\left\langle F\right\rangle}_{\mathcal{F}}\}\).
  As noted above, this also shows the result for the problem with forest
  automata as input, since the forest algebra may be converted to a
  forest automaton using only logarithmic space.
  If given a model \(\mathcal{A}\) for \(F\) we can construct \(\sigma: x\mapsto \{\mathcal{A}(x)\}\)
  which satisfies \(\sigma(L)\subseteq \mathrm{True}_{\mathcal{F}}\).
  Given a substitution \(\sigma(L)\subseteq \mathrm{True}_{\mathcal{F}}\),
  there exists a \emph{homomorphic} substitution \(\sigma'\)
  such that \(\sigma'(x)\subseteq \sigma(x)\).
  For this homomorphic substitution, the variables are substituted consistently,
  so this substitution \(\sigma'\) also defines an assignment,
  which is a model for \(F\) by the definition of \(\mathrm{True}_{\mathcal{F}}\).

  We can show \EXPTIME-hardness of~\textit{\ref{itm:existsOI:equal}} by reduction
  from the problem with a given substitution in~\cref{thm:givenOI},
  by using~\cref{lem:combineInequalities} to force the substitution to
  be the given one. Note that the resulting forest algebras or automata
  can be constructed to be of polynomial size, since the (homomorphic) substitution
  used in the hardness proof substitutes all \(x\in X\) by the same forest language.

  One way of showing that~\textit{\ref{itm:existsOI:equalboth}} is undecidable already for
  unary forest languages is
  by reduction from the problem given in~\cite[Theorem 2]{LehtinenOkhotin2010} which asks,
  given \(C,D,E,F\subseteq \mathbb{N}\), whether there exists \(X\subseteq\mathbb{N}\) such that
  \(X+X+C=X+X+D\) and \(X+E=F\).
  Identifying sets of natural numbers with subsets of \({\{a\}}^*\subseteq \mathcal{F}(A)\),
  this is the case if and only if there exists a relational substitution \(\sigma\colon\{x\}\to\mathcal{F}(A)\)
  such that \(\sigma_r(a(x+x+C)+a(x+E)) = \sigma_r(a(x+x+D)+aF)\).
  If we do not restrict the problem to unary languages, all results are generalizations of the
  respective problem for word languages~\cite{Kunc07,Okhotin2005}.
\end{proof}

\end{document}